\documentclass[graybox]{svmult}

\pdfoutput=1
\usepackage{color,multicol}

\usepackage{pstricks}

\usepackage{graphicx}
\newrgbcolor{blue}{0. 0. 0.}
\newrgbcolor{red}{0. 0. 0.}
\newrgbcolor{green}{0. 0. 0.}

\usepackage{mathptmx}       
\usepackage{helvet}         
\usepackage{courier}        
\usepackage{type1cm}        
\usepackage{makeidx}         
\usepackage{graphicx}        
\usepackage{multicol}        
\usepackage[bottom]{footmisc}

\makeindex             

\begin{document}

\title*{The LDA+DMFT route to identify good thermoelectrics}
\author{ K.\ Held, R.\ Arita, V.\ I.\ Anisimov, and K.\ Kuroki
}
\institute{K. Held \at Institute for Solid State Physics, Vienna University of Technology, A-1040 Vienna, Austria {http://www.ifp.tuwien.ac.at/cms}
\and  R.\ Arita \at Department of Aplied Physics, University of Tokyo, Tokyo 113-8656, Japan
\and K. Kuroki  \at University of Electro-Communications 1-5-1 Chofugaoka, Chofu-shi
Tokyo 182-8585, Japan  
\and 
V.\ I.\ Anisimov \at Institute of Metal Physics, Russian Academy of Science-Ural division,
620219 Yekaterinburg, Russia 
}
%
%
\newcommand{\onlinecite}{\cite}
\maketitle

\abstract{For technical applications thermoelectric  materials
with a high figure of merit are desirable, and 
strongly correlated electron systems
are very promising in this respect. Since effects of bandstructure
\textit{and} electronic correlations play an important role for
getting large figure of merits,
the combination of local density approximation \textit{and} 
dynamical mean field theory is an ideal tool for the computational materials 
design of new thermoelectrics as well as to help
us  understand the mechanisms leading
to large figures of merits in certain materials. 
This conference proceedings provides for a brief
introduction to the method and reviews recent
results for LiRh$_2$O$_4$.
}

\section{Introduction}
\label{sec:Intro}
Against the background of climate change and the present energy crisis,
the quest for alternative, green energy sources is more urgent than ever.
 In this regard, thermoelectric materials which transform waste heat (gradients) into electrical power through the Seebeck effect \cite{Seebeck,Mahan}
are particularly appealing. However, due to a low efficiency we have not
yet witnessed a wider technological application almost 200 years
after Seebeck's discovery. Instead, thermoelectrical applications
are restricted to niche markets such as radioisotope power systems for
satellites \cite{satellite}. A possible first major application
is the exhaust heat of cars and trucks,
as  automobile companies presently test thermoelectrical generators
 in prototypes \cite{cars}. Such efforts could be put on another
level if novel materials with a higher figure of merit $ZT$, where $Z$ is 
the power factor and $T$ the temperature,
and hence a higher efficiency,
 were available. Most present technical applications
use semiconductors such as Bi$_2$Te$_3$ \cite{Mahan} 
where recently power factors $Z$ considerably larger than $1$ could be achieved
 through phonon \cite{BiTePh} and bandstructure
engineering \cite{PbTeDOS}. 

Very promising are novel materials on the basis of
 strongly correlated  electron systems (SCES) \cite{Paschen} which are
at the core of the present conference proceedings. This 
class of materials is very diverse,  ranging from
metals to Kondo insulators and semiconductors, from
$d$ to $f$ electron systems, from relative simple
crystal structures such as FeSb$_2$ \cite{FeSb2} to most complex metallic cage compounds.

Having such a wide field and the additional possibilities
to nano- and heterostructure these systems, a better theoretical
understanding and reliable tools to compute thermoelectric
properties quantitatively are  mandatory. Theoretical physicists from the SCES
community have analyzed  thermoelectric
materials mainly 
on a model level, i.e., on the basis of the Falikov-Kimball, 
Hubbard and periodic Anderson model \cite{TEPAM1,TEPAM2}, often
employing dynamical mean field theory (DMFT) \cite{DMFT1,DMFT2,DMFT3}.
These calculations showed, among others,  the importance of
correlation-induced enhancements of the effective mass generating
a high, but narrow density of states --or spectral function
to be precise-- close to --but not at--  the Fermi level. As a consequence,
 the thermoelectric figure of merits can be strongly enhanced.
On the other hand,  theoreticians from the density functional theory (DFT) \cite{DFT} community have been emphasizing
the importance of a particularly  high
density of states (DOS) \cite{Singh,Wilson}  and of the 
large group velovities for certain shapes of the bandstructure \cite{Kuroki}.

Since both, correlations and bandstructure, can substantially contribute
to enhanced thermoelectrical figures of merit, we need to deal with
both of them on an equal footing. Only if both aspects are optimized
we can expect to design  materials or artificial heterostructures
with a really large figure of merits.  
Taking correlations and bandstructure into account is possible with the
merger  \cite{LDADMFT0,LDADMFT1} of DFT in its local density approximation (LDA) \cite{LDA} and DMFT,
for which the name LDA+DMFT was coined \cite{Nekrasov00}, see \onlinecite{LDADMFT2,LDADMFT3,LDADMFT4} for reviews. While LDA+DMFT  has been applied already
to many SCES materials, thermoelectrical properties have been calculated
rarely in the past. Noteworthy exceptions are
LaTiO$_3$  \cite{Oudovenko} and LiRh$_2$O$_4$ \cite{Arita08}.
The main reasons for this is that a wider experimental interest
in SCES thermoelectrics emerged rather recently and that
the calculation of thermoelectric properties such as the
Seebeck coefficient requires some additional postprocessing
which is not yet  standard in LDA+DMFT calculations.

\subsection{Outline}

In the following, we will give a brief, elementary introduction
to the LDA+DMFT approach in Sec. \ref{sec:LDADMFT}. This Section
is divided into the three  steps LDA (Sec.\ \ref{sec:LDA}),
DMFT  (Sec.\ \ref{sec:DMFT}), and the necessary postprocessing
for calculating thermoelectrical response functions
(Sec.\ \ref{sec:Post}).  Sec. \ref{sec:LiRh2O4}
presents exemplary results by hands of LiRh$_2$O$_4$
which are reproduced from Ref.\  \onlinecite{Arita08}.
Finally,  Sec. \ref{sec:Outlook} gives a summary and
an outlook.

\section{LDA+DMFT method}
\label{sec:LDADMFT}
The aim of this section is to give the reader a brief, elementary introduction
to the LDA+DMFT approach; for more details see the reviews
\cite{LDADMFT2,LDADMFT3,LDADMFT4}.

Starting point  is the general {\em ab-initio} Hamiltonian for
every material which, without relativistic corrections, reads
in the Born-Oppenheimer approximation
\vspace{.395cm}

\noindent \hspace{1.3cm} \textcolor{blue}{kinetic energy}  \hspace{.9cm} \textcolor{green}{ lattice potential} \hspace{1.4cm} \textcolor{red}{ Coulomb interaction}
\vspace{.395cm}

\noindent \hspace{1.699cm} \includegraphics[clip=true,height=1.6cm]{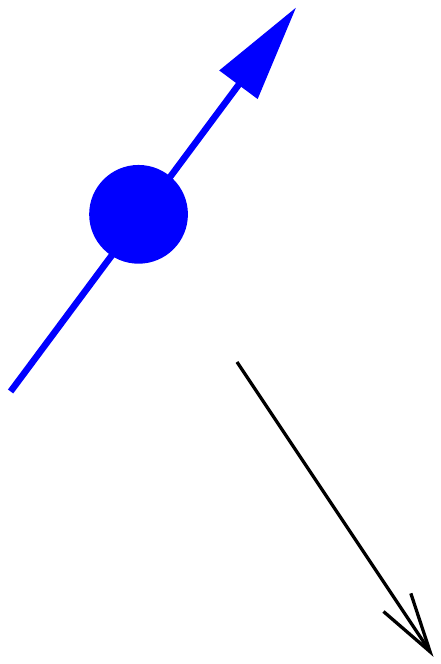}
\hspace{1.2979cm}  \includegraphics[clip=true,height=1.6cm]{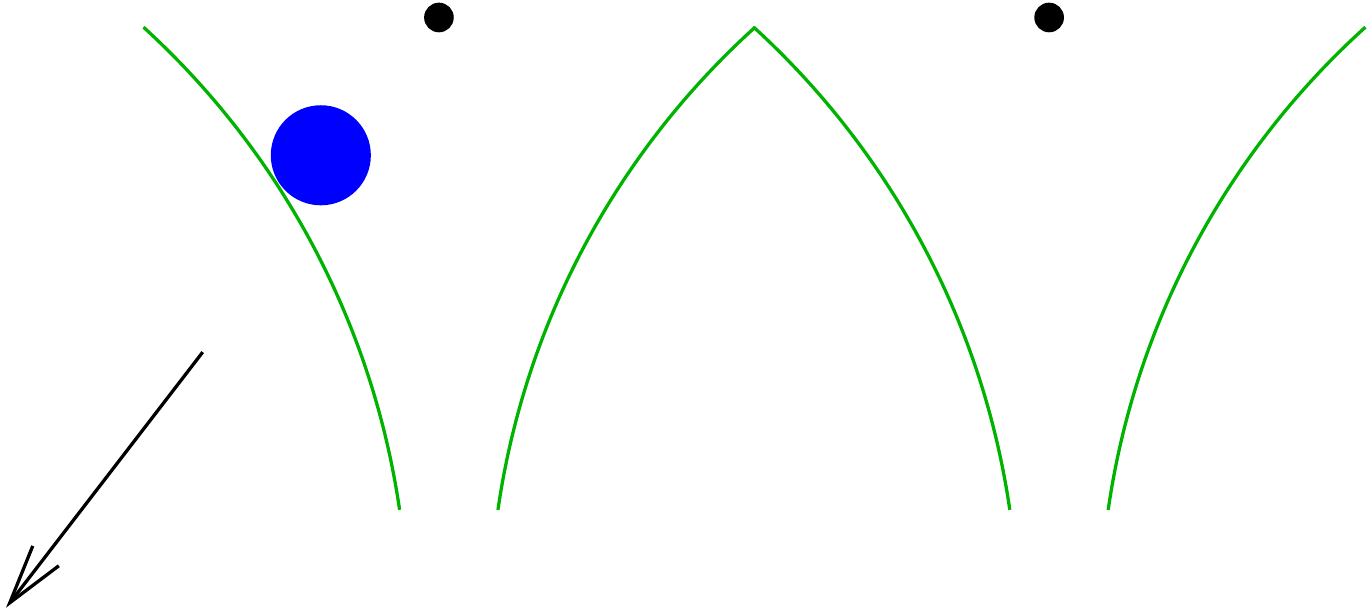}
\hspace{.5cm}  \includegraphics[clip=true,height=1.6cm]{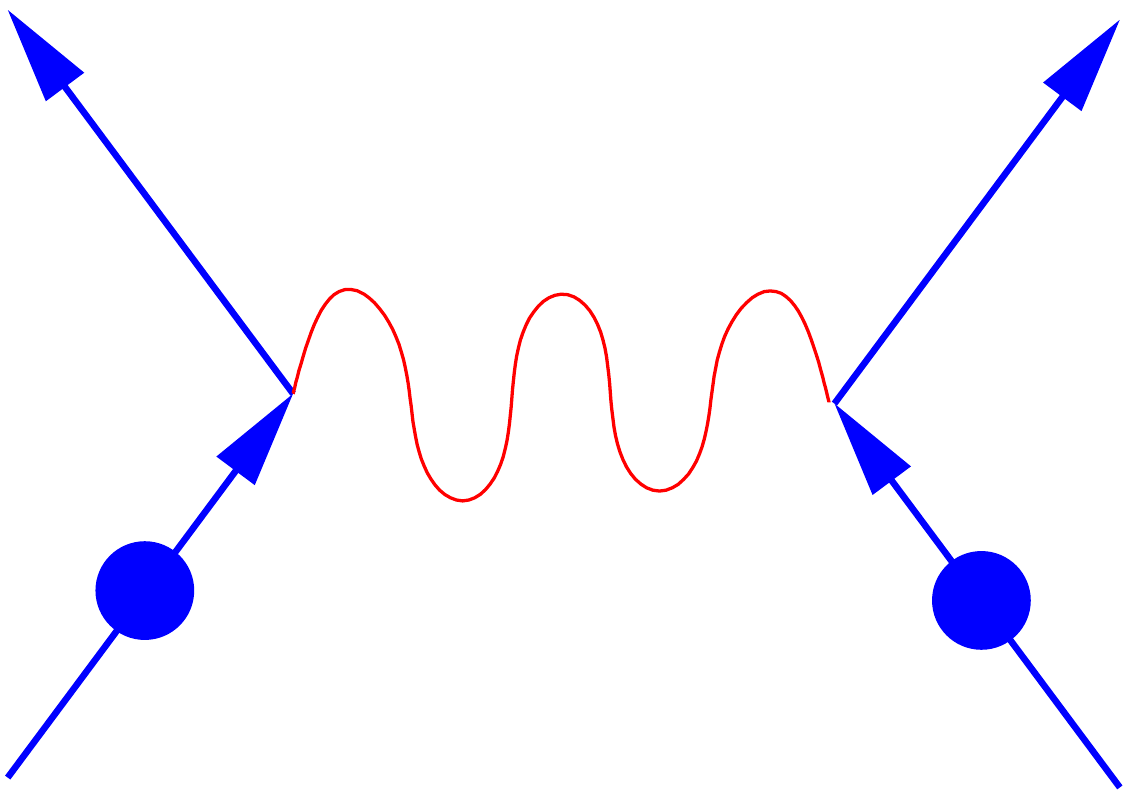}
\vspace{-.295cm}

\begin{equation}
 {H} = \sum_i  
     \left[ \;\textcolor{blue}{ -\frac{\displaystyle  \hbar^2 \Delta_i }{\displaystyle 2 m_e}}\;+\; \textcolor{green}{\sum_l \frac{\displaystyle - e^2}{\displaystyle 4\pi\epsilon_0} \; \frac{\displaystyle Z_l}{\displaystyle |{\bf r}_i-{\bf R}_l|}}  \;\right]\;\;\; +\;\;\;
    \frac{1}{2} \; \sum_{i\neq j} \; \textcolor{red}{\frac{\displaystyle  e^2}{\displaystyle 4\pi\epsilon_0}\; \frac{\displaystyle 1}{\displaystyle |{\bf r}_i -{\bf r}_j|}}
\label{Hamabinitio}
\end{equation}
It consists of three terms: 
1.) The kinetic energy
given by the Laplace operator 
$\Delta_i$, Planck constant $\hbar$, and mass $m_e$ for every electron $i$. 
2.) The lattice potential given by the Coulomb interaction
between (static) ions at position ${\bf R}_l$ with charge $Z_l e$
 and electrons at position ${\bf r}_i$ with charge $-e$.
3.) Finally, the Coulomb interaction between each pair of
electrons $i$ and $j$ [note the factor 1/2 is needed since each pair is counted twice in Eq.\ (\ref{Hamabinitio})]. Input for the LDA+DMFT calculation
is usually the experimental crystal structure, i.e., the positions $R_i$ as
an adequate relaxation procedure to determine the  $R_i$'s from theory still needs to be developed.

While Hamiltonian  (\ref{Hamabinitio}) is easy to write down, it is impossible
to solve, even numerically, for more than ${\cal O}(10)$ electrons, since the
 movement of every electron is {\em correlated} with that of
every other electron through the last term: the Coulomb interaction 
between the electrons.  These electronic correlations play a particularly
important role if electrons are confined in or $d$ or $f$-electrons
or in artificial nanostructures. For such systems the typical distance
$|{\bf r}_i -{\bf r}_j|$ 
between two such electrons on the same lattice site (i.e., two electrons
in the set of $d$- or $f$-orbitals around the same ion) is small
so that the Coulomb interaction and, hence, also the electronic correlations are strong.

\subsection{LDA step}
\label{sec:LDA}
Since it is impossible to solve Hamiltonian  (\ref{Hamabinitio}), we have to
develop approximations, and arguably the most successful approximation
so far are those developed within the DFT framework,
particularly the LDA \cite{LDA}.
Strictly speaking, DFT only allows to calculate ground state
energies and its derivatives but {\em not} bandstructures and thermoelectric
transport functions. However, it turned out that the auxiliary
Kohn-Sham Lagrange parameters ${\mathbf \epsilon}_k$ often also
describe bandstructures very accurately, making bandstructure
calculations one of the major applications of LDA.
 Interpreting 
the LDA Lagrange parameters ${\mathbf \epsilon}_k$
as the physical (one-electron) excitation energies, i.e.,
the bandstructure, corresponds to replace 
 Hamiltonian  (\ref{Hamabinitio}) by the Kohn-Sham
\cite{KohnSham} LDA Hamiltonian
\begin{equation} {H}_{\rm LDA}\! =\! \sum_i
    \! \left[\!\textcolor{blue}{-\frac{\displaystyle \hbar^2 \Delta_i }{\displaystyle 2 m_e}}
 +\! \textcolor{green}{\sum_l \frac{\displaystyle -e^2}{\displaystyle 4\pi\epsilon_0} \frac{\displaystyle 1}{\displaystyle |{\bf r}_i-{\bf R}_l|}} +\! \textcolor{red}{
\int \! {\rm d}^3 r \,\frac{\displaystyle e^2}{\displaystyle 4\pi\epsilon_0} \frac{\displaystyle 1}{\displaystyle |{\bf r}_i-{\bf r}|} \rho ({\bf r})} 
+ {\textcolor{red}{{ {V_{xc}^{\rm{LDA}}(\rho({\bf r}_i))}}
}}\!\right]\!
\label{HLDA} 
\end{equation}
This Hamiltonian shows that the complicated electron-electron interaction
causing the complicated electronic correlations has been
replaced by two simpler terms: The Hartree term describing
the Coulomb interaction of electron ${\bf r}_i$ with the time-averaged 
mean density $\rho({\bf r})$ of all electrons and an additional term
${\textcolor{red}{{ {V_{xc}^{\rm{LDA}}}}}}$ which aims at including
the effects of correlations and interactions. 

However, the exact
form of this term is unknown and certainly it is not local in
${\bf r}$ as approximated in the LDA. One can
take the ${\textcolor{red}{{ {V_{xc}}}}}$ of the jellium model
\cite{jellium} which has a constant electron density and is only
weakly correlated. Hence, it is not surprising that
LDA bandstructure calculations fail for SCES \cite{LDA}. For such materials,
which are at the focus here, we need to take electronic correlations
into account more profoundly. 

A possibility to do so is to take the LDA bandstructure of
the less correlated orbitals but to supplement that of the more
correlated $d$- or $f$-orbitals by explicitly taking into account
the most, important local Coulomb interaction.
This leads to the Hamiltonian
\begin{eqnarray}
\hat{\cal{H}} =\underbrace{\sum_{{\bf k} l m\sigma} \textcolor{green}{\epsilon^{\rm LDA}_{{\bf k} lm}} \hat{c}^{\dagger}_{{\bf k} l \sigma} \hat{c}^{\phantom{\dagger}}_{{\bf k} m \sigma}}_{\displaystyle H_{\rm LDA}} + \textcolor{red}{\frac{\displaystyle 1}{2}\! \sum_{i\,  l\sigma m\sigma'}\!\!\!
\textcolor{red}{U_{l m}^{\sigma \sigma'}}\; \hat{n}_{i  l \sigma\phantom{'}\!\!}\,\hat{n}_{i\phantom{l}\!\! m \sigma{'}}} -
 \textcolor{red}{{\Delta \epsilon} \sum_{i m\sigma} \hat{n}_{i m \sigma}}
\label{HamLDAU},
\end{eqnarray}
where the first part is the same as the LDA Hamiltonian
(\ref{HLDA}) but in second (instead of first) quantization
and in ${\bf k}$ and orbital space  (with $l$ and $m$ denoting two different orbitals) with creation
and annihilation operators 
$\hat{c}^{\dagger}_{{\bf k} l \sigma}$ and $\hat{c}^{\phantom{\dagger}}_{{\bf k} m \sigma}$, respectively. 

\begin{figure}[tb]
\centerline{\includegraphics[clip=false,width=5.7cm]{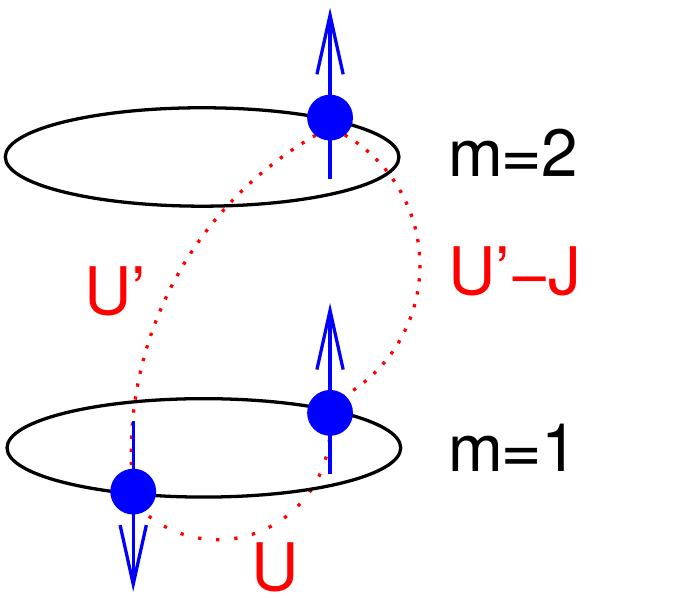}}
\caption{\label{Fig:Hund}
Illustration of the different elements of the Coulomb interaction matrix
of Hamiltonian (\ref{HamLDAU}). There is an inter-orbital Coulomb repulsion $U'$, which is reduced  by Hund's exchange $J$ for a ferromagnetic spin alignment,
and an  intra-orbital interaction $U$. Orbital rotational symmetry relates
these quantities as $U=U'+2J$.
}
\end{figure}

The second term explicitly takes the local
Coulomb interaction on the same ion site $i$ into account. Typically only the
Coulomb interactions for  $d$ (or $f$) $l$ and $m$ orbitals are considered
here. These interactions
are spin and orbital dependent because of the
the exchange matrix elements leading to Hund's rules, see Fig.\ \ref{Fig:Hund} for an illustration.
Let us note that in Hamiltonian
(\ref{HamLDAU}) only the density-density terms are included
since the inclusion of the spin-flip terms of Hund's exchange
became only possible in quantum Monte Carlo (QMC) simulations \cite{Hirsch} with recent improvements  \cite{Rubtsov04a,Rubtsov05a,Werner05a,Werner06a,Sakai}.

Finally the third $\Delta \epsilon$ term subtracts those contributions of $U$ already taken into account
in the LDA to avoid a double counting. For a truly {\rm ab-initio} 
calculation,
$U'$, $J$, and $\Delta \epsilon$ still need to be determined.
To this end, screening has to be taken into
account; and a possibility within the LDA framework is to employ
constrained LDA, for details see \cite{LDADMFT4}.

\subsection{DMFT step}
\label{sec:DMFT}
Having derived a multi-orbital many-body Hamiltonian
(\ref{HamLDAU}) from the {\em ab-initio} Hamiltonian (\ref{Hamabinitio}),
we still need to solve it. A possible way to do
so is to use Hartree-Fock, allowing for symmetry breaking
with respect to the spin and orbital elements, i.e.,
\begin{equation}
\frac{\displaystyle 1}{2} \sum_{i\,  l\sigma m\sigma'}\!\!\!
\textcolor{red}{U_{l m}^{\sigma \sigma'}}\; \hat{n}_{i  l \sigma\phantom{'}\!\!}\,\hat{n}_{i\phantom{l}\!\! m \sigma{'}} \rightarrow
 \sum_{i\,  l\sigma m\sigma'}\!\!\!
\textcolor{red}{U_{l m}^{\sigma \sigma'}}\; \hat{n}_{i  l \sigma\phantom{'}\!\!}\,\langle \hat{n}_{i\phantom{l}\!\! m \sigma{'}}\rangle - \frac{\displaystyle 1}{2}  \sum_{i\,  l\sigma m\sigma'}\!\!\!
\textcolor{red}{U_{l m}^{\sigma \sigma'}}\;\langle \hat{n}_{i  l \sigma\phantom{'}\!\!}\rangle \,\langle \hat{n}_{i\phantom{l}\!\! m \sigma{'}}\rangle,
\label{Eq:HF}
\end{equation}
where $\langle \hat{n}_{i\phantom{l}\!\! m \sigma{'}}\rangle$
is the average occupation of the orbital $m$ on site $i$ with
spin $\sigma{'}$. However, in this LDA+$U$  \cite{LDAU} approach
electronic correlations are neglected through Eq.\ (\ref{Eq:HF});
and the only chance to reduce the Coulomb interaction energy is
by a strong symmetry breaking. Hence, tendencies to 
magnetic or orbitally ordered phases are grossly overestimated,
as is the tendency to open gaps. Even within these ordered phases
many-body aspects such as spin-polarons are neglected as was shown in
\cite{spinpolaron}.

A reliable approximation to include the {\em local} correlations
induced by the {\em local} Coulomb interaction of Hamiltonian (\ref{HamLDAU})
is possible with DMFT \cite{DMFT1,DMFT2,DMFT3}. We cannot derive this approach
in full detail here and refer the interested
reader to \cite{DMFT2} and \cite{LDADMFT4}.
The basic idea is visualized in Fig.\ \ref{Fig:DMFT}:
\begin{figure}[tb]
\includegraphics[clip=false,width=3.7cm,angle=270]{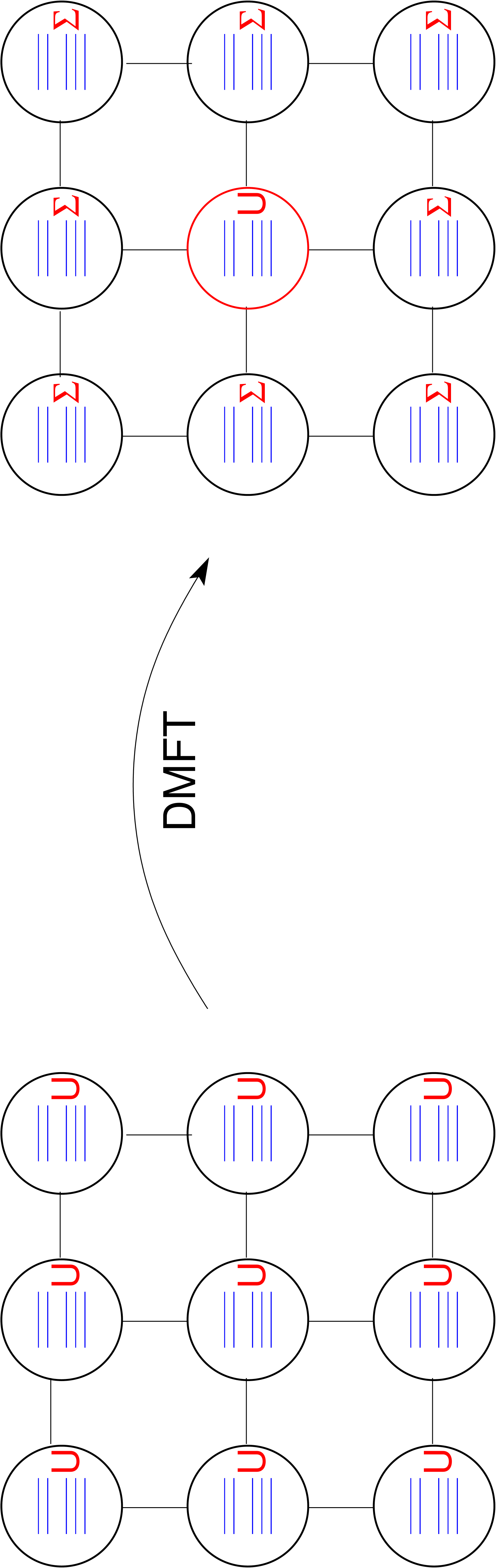}

\caption{\label{Fig:DMFT}
In DMFT, we approximate the material specific lattice Hamiltonian
(\ref{HamLDAU}) by a problem where the interaction is replaced by a self energy on all sites except for  one. This DMFT single-site problem is equivalent
to an Anderson impurity model which has to be solved self-consistently
together with the ${\mathbf k}$-integrated Dyson eq. (\ref{Eq:Dyson}). 
}
\end{figure}
We replace the local interaction on all sites but one by a self-energy 
$\Sigma(\omega)$.
This gives rise to an Anderson impurity model of a single interacting 
site in a medium  ${\cal G}_0(\omega)$
given by the self energy and the interacting Green function $G(\omega)$:
\begin{equation}
{\cal G}_0(\omega)^{-1} = G(\omega)^{-1}+\Sigma(\omega)
\end{equation}
This Anderson impurity model, defined by its non-interacting Green function
${\cal G}_0$ has to be solved self-consistently
together with the ${\mathbf k}$-integrated Dyson equation,
where the LDA bandstructure $\epsilon^{\rm LDA}_{l,m}({\mathbf k})$ enters
as a matrix in the orbital indices ($V_{{\rm BZ}}$ denotes the volume of the Brillouin zone):
\begin{equation}
\textcolor{blue}{G}^\sigma_{lm}(\omega) = \!\int \!  \frac{{\rm d}^3 k}{V_{{\rm BZ}}}
\left[\omega\!+\!\mu\!-\![\textcolor{green}{\epsilon^{\textcolor{green}{\rm LDA}}}-\textcolor{red}{\Delta \epsilon}]_{lm}( {\bf k})
\!\!- \!\!\textcolor{blue}{\Sigma}^\sigma_{lm}(\omega)\right]^{-1}.
\label{Eq:Dyson}
\end{equation}

From a diagrammatic point of view, DMFT corresponds to all
(topologically distinct) Feynman diagrams of which, however,
only the local contribution for the self energy is taken into account.
Hence, it is non-perturbative in the Coulomb interaction but neglects
non-local correlations between sites. Recent improvements of DMFT
include such non-local correlations by taking a cluster of interacting sites instead of a single one in Fig.\ \ref{Fig:DMFT} \cite{clusterDMFT1,clusterDMFT2,clusterDMFT3}
or by extending the diagrammatic contributions in the
dynamical vertex approximation (D$\Gamma$A) \cite{DGA1}, also see 
\cite{DGA2,DGA3,DGA4}.

What we still need to do is to solve the Anderson impurity model self-consistently,
which for realistic multi-orbital calculations is typically done
by quantum Monte Carlo simulations, different approaches are discussed
in \cite{LDADMFT4}.
The standard result of such a DMFT(QMC) calculation is the interacting local
Green function  $\textcolor{blue}{G}(i \omega_{\nu})$
for imaginary (Matsubara) frequencies $i \omega_{\nu}$ or its
Fourier transform, the imaginary time Green function
 $\textcolor{blue}{G}(\tau)$. But also various correlation
functions and susceptibilities can be calculated.

\subsection{Calculation of thermoelectrical response functions}
\label{sec:Post}
Starting point for calculating transport properties is the
Kubo formula. For thermoelectric materials the Seebeck coefficient 
\begin{equation}
\textcolor{red}{S}= - \frac{k_{\rm B}}{|e|}\frac{\textcolor{blue}{A_1}}{\textcolor{blue}{A_0}}
\end{equation}
is of particular importance. It is given by the constants
Boltzmann   $k_B$, unit charge $e$  and
  the ratio
of two correlation functions, the current--current and the
current--heat-current correlation function
\begin{eqnarray}
{\textcolor{blue}{A_0}}&=& \lim_{i\nu\rightarrow 0}
\frac{i\hbar k_B T} {i\nu } \int_0^{\beta} {\rm d}\tau\, e^{i\nu\tau}\, \langle T_\tau \textcolor{blue}{j}(\tau) \textcolor{blue}{j}(0) \rangle\label{Eq:A0}
\\
{\textcolor{blue}{A_1}}&=& \lim_{i\nu\rightarrow 0}
\frac{i\hbar}{i\nu} \int_0^\beta {\rm d}\tau \, e^{i\nu\tau}\, \langle T_\tau 
\textcolor{blue}{j}(\tau) \textcolor{blue}{j_Q}(0) \rangle
\label{Eq:A1}
\end{eqnarray}
in the static limit, i.e., frequency $i\nu\rightarrow 0$.
Here,  $T_\tau$ is Wick's time ordering operator;
  $\textcolor{blue}{j}(\tau)$ and  $\textcolor{blue}{j_Q}(\tau)$ 
are the current and heat-current operators respectively.
Also relevant is the heat-current--heat-current correlation
function 
\begin{equation}
{\textcolor{blue}{A_2}} = \lim_{i\nu\rightarrow 0}
\frac{i\hbar}{i\nu k_B T} \int_0^\beta {\rm d}\tau \, e^{i\nu\tau}\, \langle T_\tau 
\textcolor{blue}{j_Q}(\tau) \textcolor{blue}{j_Q}(0) \rangle
\label{Eq:A2}
\end{equation}
which yields the electronic contribution to the thermal conductivity $\kappa$
similar as $A_0$ does for the electrical conductivity $\sigma$.
Since the phononic contribution to the thermal conductivity
is however typically much larger at room temperature and can be reduced by phonon engineering, we will 
not consider $\kappa$  in the following. Instead we will
concentrate of the purely electronic contributions to the power
factor
\begin{equation}
Z= \frac{S^2 \sigma}{\kappa},
\label{seebeckeq1}
\end{equation}
i.e., on $S$ and $\sigma$.

\begin{figure}[tb]

\centerline{\includegraphics[clip=true,width=0.4\textwidth]{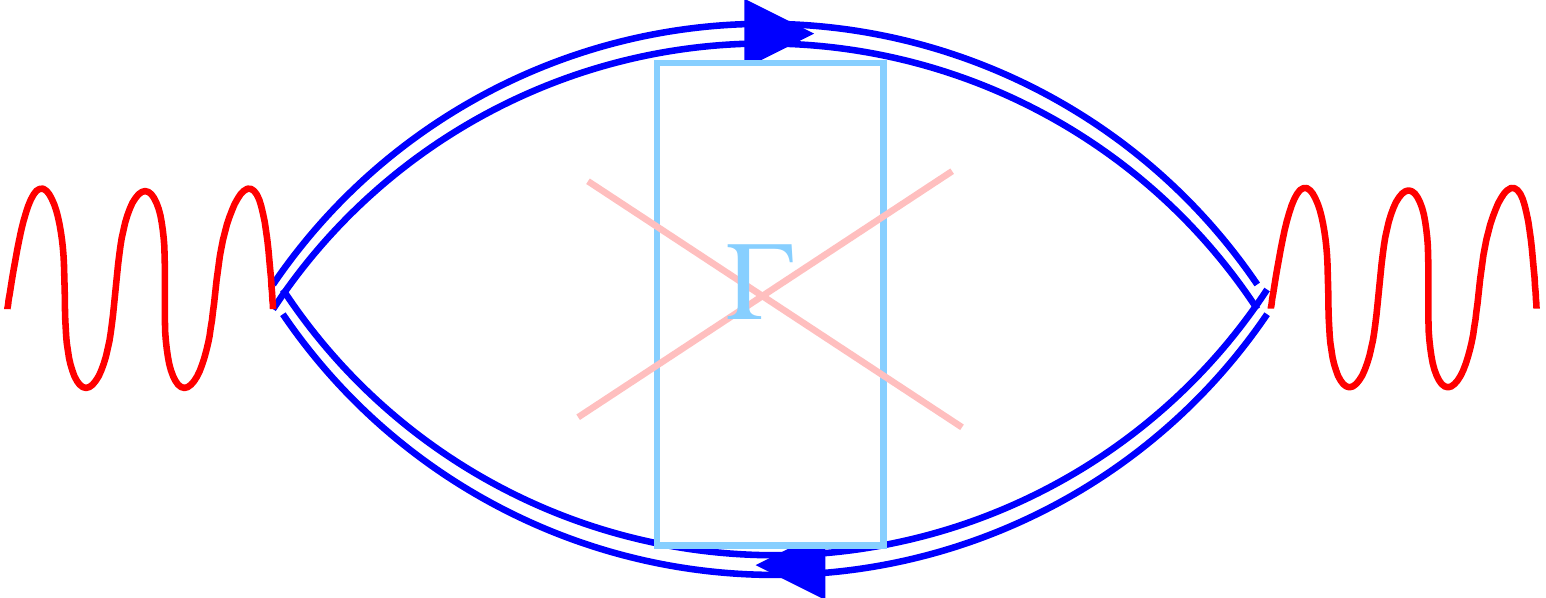}}
\caption{\label{Fig:Bubble}
Diagrammatic representation of the (heat-)current--(heat-)current
 correlation functions Eqs. (\ref{Eq:A0},\ref{Eq:A1},\ref{Eq:A2}) with an incoming
frequency (wiggled line) $i\nu\rightarrow 0$. 
For the current operator the wiggled line yields a factor
$\textcolor{red}{v^x({\bf k})}$, for the heat-current
operator  a factor
$\epsilon^{\rm LDA}_{\bf k} \textcolor{red}{v^x({\bf k})}$.
The vertex  $\Gamma$ 
is typically neglected as indicated so that the calculation of the correlation
function reduces to the simple bubble diagram of two (interacting) 
Green function, i.e., a factor $\textcolor{blue}{G(k,\omega)}$
for each of the two double lines.
}
\end{figure}

Diagrammatically, the correlation functions Eqs.\ (\ref{Eq:A0},\ref{Eq:A1},\ref{Eq:A2})
correspond to Fig.\ \ref{Fig:Bubble}. As indicated, the vertex $\Gamma$ is usually not
taken into account. In  case of full orbital degeneracy (of the low energy orbitals), this
holds exactly since one
can show by a simple argument that vertex contributions are, for the local DMFT vertex, odd
in ${\mathbf k}$ and hence their integrated contribution vanishes, see
 \cite{Pruschke}. In the case of LiRh$_2$O$_4$ where the three low energy orbitals
are  very similar in energy and occupation, neglecting the vertex is still justified
approximately. 
This allows us to calculate the bubble diagram for the (heat-)current--(heat-)current correlation functions $A_m$ from the spectral function $\rho({\bf k},\omega)=-1/\pi\;{\rm Im}G({\bf k},\omega)$. In the $x$-direction, we obtain for
diagram Fig.\ \ref{Fig:Bubble}
\begin{eqnarray*}
\textcolor{blue}{A_m}&=&2\pi\hbar \int^{\infty}_{-\infty}d\omega \;
\frac{1}{V}\sum_{\bf k} {\rm Tr} \big[
\textcolor{red}{{v^x({\bf k})}} \textcolor{blue}{\rho({\bf k},\omega)}
\textcolor{red}{v^x({\bf k})}\textcolor{blue}{\rho({\bf k},\omega)}\big] \; f(\omega)f(-\omega)\textcolor{red}{(\beta \omega)^m}.
\label{seebeckeq2}
\end{eqnarray*}
Here $v_{\mathbf k}$ are in the general formalism the dipole matrix
elements which we replaced approximately  by the simpler group  
velocity obtained through the derivative of the dispersion relation:
\begin{equation}
\textcolor{red}{v_{\mathbf k}}=\textcolor{red}{\frac{\partial \epsilon^{\rm LDA}_{\mathbf k}}{\partial {\mathbf k}}}.
\end{equation}
These are relatively easy to calculate from the LDA bandstructure.
Note, here the quantities $\textcolor{red}{v_{\mathbf k}}$, 
 $\textcolor{red}{\epsilon^{\rm LDA}_{\mathbf k}}$, and  $\rho({\bf k},\omega)$
are all matrices in the orbital indices.

What we still need for calculating the DMFT (heat-)current--(heat-)current correlation functions, is the ${\mathbf k}$-dependence of  $\rho({\bf k},\omega)$.
In the DMFT self-consistency cycle, one calculates however only the local Green function 
\begin{equation}
G(i\omega_{\nu})=\frac{1}{V}\sum_k G({\bf k},i\omega_{\nu})
\end{equation}
at Matsubara frequencies $i \omega_{\nu}$.
From the DMFT $G(i \omega_{\nu})$ or 
its Fourier transform the imaginary time
$G(\tau)$, we can  determine the optical and thermal conductivity as well as the Seebeck coefficient in some post processing steps:

First, we need the self-energy for (real) frequencies.
The standard procedure \cite{Nekrasov} to this end is first to
analytically continue the Green function to real frequencies.
This is done by the maximum entropy method \cite{MaxEnt}
yielding ${\rm Im} G(\omega)$ at real frequencies $\omega$ from $G(\tau)$.
From this, in a second step, the full  $G(\omega)$ is 
constructed by Kramers-Kronig transformation.
Third, that self energy is determined which, if plugged into the
 ${\mathbf k}$-integrated Dyson eq. (\ref{Eq:Dyson}), gives the Green
function which is closest to the QMC-determined  $G(\omega)$.
Finally from the self energy $\Sigma(\omega)$ for real frequencies,
we can determine 
\begin{eqnarray*}
\textcolor{blue}{G}^\sigma_{lm}(\omega) = 
\left[\omega\!+\!\mu\!-\![\textcolor{green}{\epsilon^{\textcolor{green}{\rm LDA}}}-\textcolor{red}{\Delta \epsilon}]_{lm}( {\bf k})
\!\!- \!\!\textcolor{blue}{\Sigma}^\sigma_{lm}(\omega)\right]^{-1},
\end{eqnarray*}
or its imaginary part   $\rho({\bf k},\omega)$.

For the LiRh$_2$O$_4$ calculations presented below it turned out that this
standard approach does not work so well because the high spectral weight close
to the Fermi level makes the analytical calculation very sensitive to
the statistical QMC error. On the other hand, we only need $\textcolor{blue}{\Sigma}(\omega)$ at small frequencies, of the order $k_{\rm B} T$, if we are interested in thermodynamical responses to at static fields (as $S$ and $\sigma$). 
Hence in \cite{Arita08}, we did 
a Pad\'e  fit \cite{Pade} to $\Sigma(i \omega_{\nu})$ which works rather
well for not too large frequencies. Comparing it with a polynomial fit
allowed us to estimate the error in  $\textcolor{blue}{\Sigma}(\omega)$,
see Fig.\ \ref{Fig:Sigma} in Sec. \ref{sec:LiRh2O4} below.

Let us note the connection to the Boltzmann approach. This is obtained
for non-interacting electrons and a constant-$\tau$ approximation,
i.e., calculation Eq.\ (\ref{seebeckeq2}) for a self
energy $\Sigma(\omega)=-i/\tau$. This reduces Eq.\ (\ref{seebeckeq2}) to
\begin{eqnarray}
\textcolor{blue}{A_m}&=&\sum_k \tau \; \textcolor{red}{v^x({\bf k})}
\textcolor{red}{v^x({\bf k})}
\left[-\frac{\partial f(\epsilon)}{\partial \epsilon}\right] \left(\frac{\epsilon^{\rm LDA}({\bf k})}{k_B T}\right)^m.
\label{Eq:Boltzmann}
\end{eqnarray}
Note that, in contrast to the thermal and electrical conductivity,  $\tau$ cancels in the Seebeck coefficient since we divide
$A_1$ by $A_0$. Hence, the exact value of the difficult to determine relaxation
time is not relevant, as long as it is constant.

For a better understanding of the microscopic origin of a large thermopower, at least as far as trhwe bandstructure effects are concerned, we can approximate the Boltzmann Eq.\ (\ref{Eq:Boltzmann}) by summing only
 the states in a window $\pm k_{\rm B}T$ around the Fermi energy (indicated by the tilde below):
\begin{eqnarray}
\textcolor{blue}{A_0} \approx \textcolor{blue}{\tau}\tilde{\sum}_k \textcolor{green} {v_A^2}+\textcolor{red}{v_B^2}\;
\label{Eq:BA0} ; &&
\textcolor{blue}{A_1}\approx \textcolor{blue}{\tau}\tilde{\sum}_k \textcolor{green}{v_A^2}-\textcolor{red}{v_B^2}.
\label{Eq:BA1}
\end{eqnarray}
Here ${v_A^2}$ and  ${v_B^2}$ are the typical (averaged) velocities above
and below the Fermi level, respectively. For the current--current correlation function $A_0$ these two contributes have to be added, whereas they have to 
be subtracted for the heat-current--current correlation function $A_1$.
The reason for the latter  is that a quasiparticle above the Fermi level carries a positive energy contribution relative to the Fermi energy, while we have a negative energy-contribution for  quasi-hole excitations below the Fermi level. For getting an (absolutely) large
Seebeck coefficient we need a large $A_1$ relative to $A_0$. Since
$A_1$ is the difference of the same (positive) contributions which are added 
 $A_0$, this requires the minuend to be much smaller than the
 subtrahend in Eq.\ (\ref{Eq:BA1}) or vice versa.
This is possibly if either (i) there are many more states below the
Fermi level than above (or vice versa for a large negative $S$) or (ii)
 the group velocity ${v_A^2}$ above the Fermi level 
is much larger than  ${v_B^2}$ (or vice versa).
Optimal would indeed be a combination of both. The route (i) can be heavily effected by electronic correlations, e.g., if we have a sharp Kondo peak directly above or below the Fermi level, but also bandstructure effects play a role.
In contrast for mechanism (ii) the LDA group velecities (or dipole matrix elements) enter so that electronic correlations are not direclty relevant.

\section{An example: LiRh$_2$O$_4$}
\label{sec:LiRh2O4}
\begin{figure}[tb]
\begin{center}
\includegraphics[width=6.5cm]{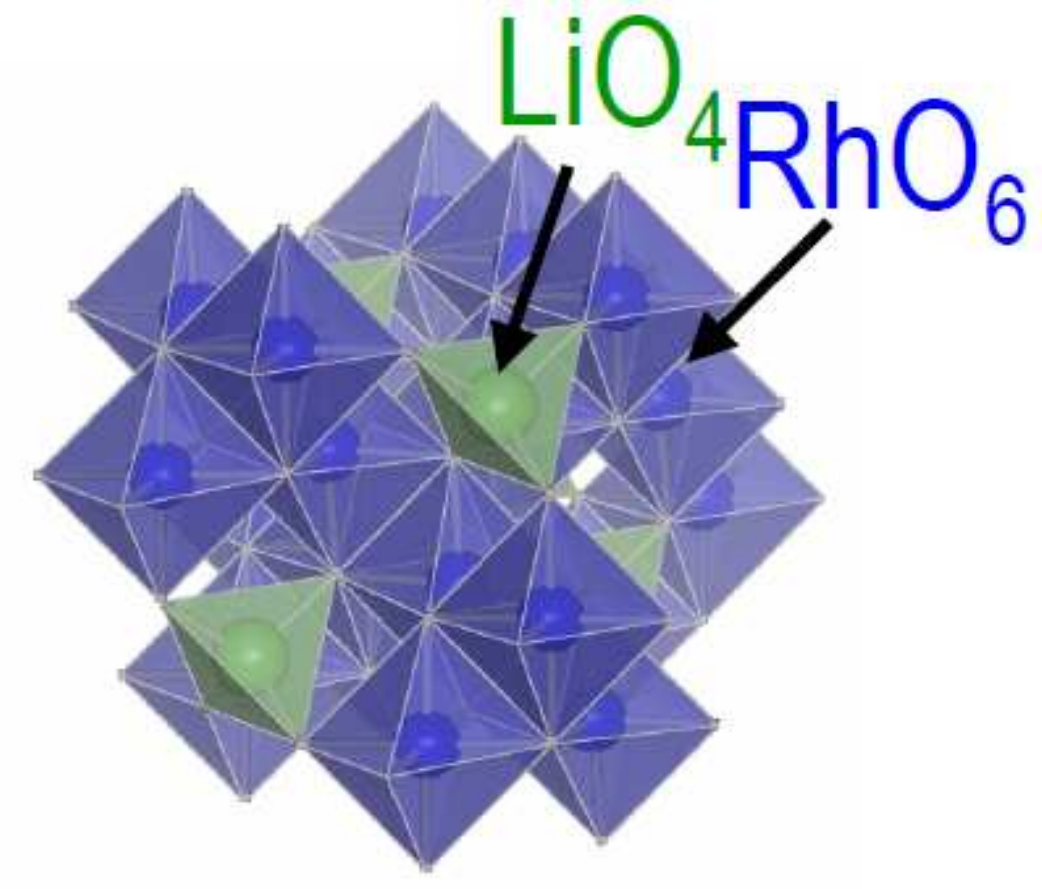}
\end{center}
\caption{{Crystal structure of  LiRh$_2$O$_4$, made up from
LiO$_4$ tetrahedra and the RhO$_5$ octahedra [reproduced from  \cite{Okamoto}].
 }}
\label{Fig:Crystal}
\end{figure}

Let us here briefly review the calculation of the Seebeck
coefficient for  LiRh$_2$O$_4$ of Ref. \cite{Arita08}.
This mixed-valent spinel, see Fig.\ \ref{Fig:Crystal}, 
was most recently synthesized by 
Okamoto {\it et al.} \cite{Okamoto}.
It shows  two structural phase 
transitions:  cubic-to-tetragonal transition 
at 230K and  tetragonal-to-orthorhombic transition at
170K. For the high-temperature cubic phase Okamoto {\it et al.}
reported a 
thermopower as large as $80 \mu$V/K at 800K, which for a metallic
system, is quite exceptional. Together with Na$_x$CoO$_2$ \cite{Terasaki},
it shows that transition-metal-oxides are promising candidates for
thermoelectric application since,
even in the metallic phase, large power factors ($S^2 \sigma$) are possible.
Conerning Na$_x$CoO$_2$, these experimental findings led to some ``heated discussion''
on the origin of the large Seebeck coefficient  on the theoretical side
\cite{Koshibae,Kuroki,hotdebate}. This makes a reliable {\em ab-initio} calculation which can put the theoretical ideas on a more solid fundament mandatory.

\begin{figure}[tb]
\begin{center}
\includegraphics[width=10.5cm]{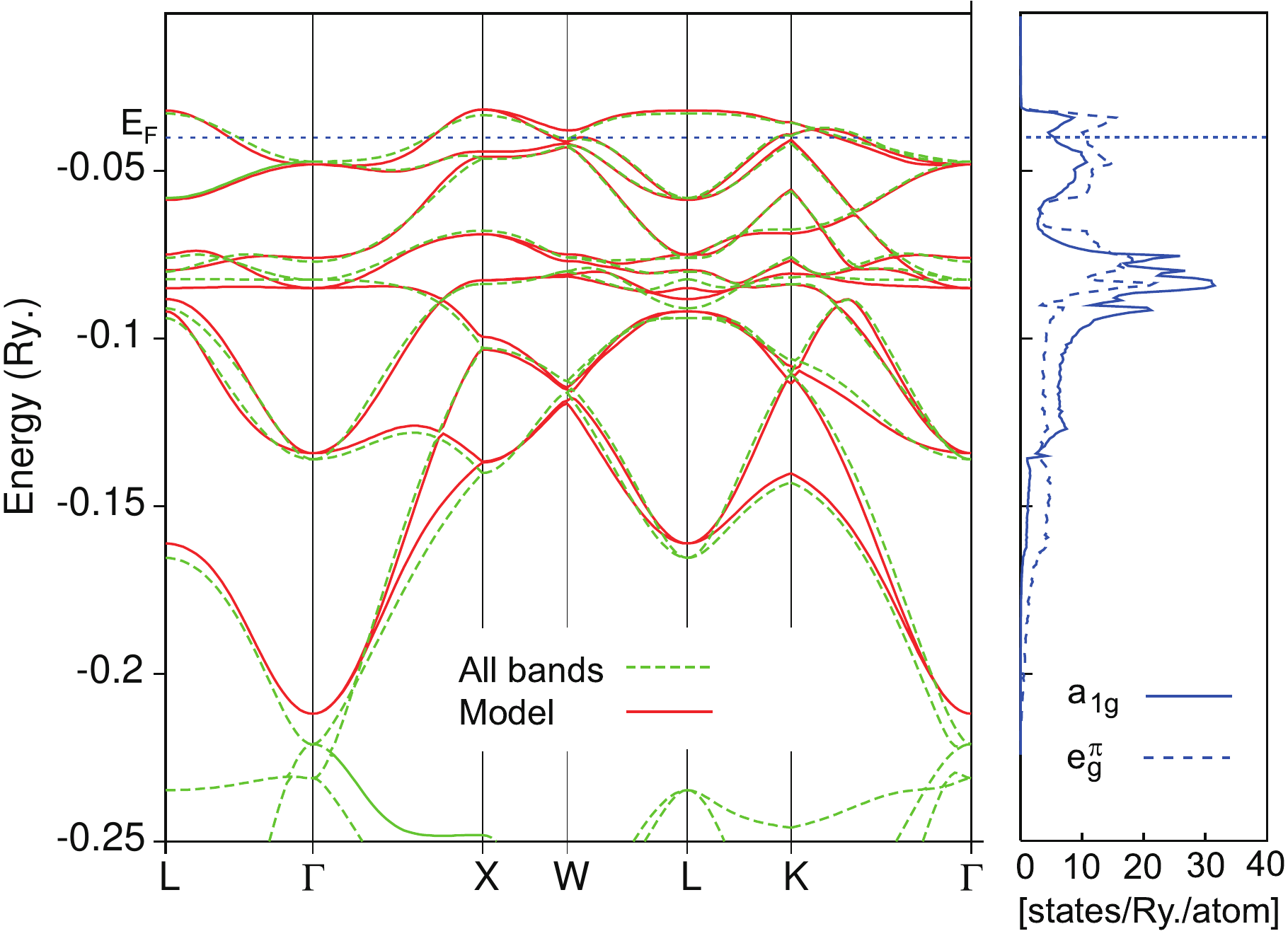}
\end{center}
\caption{{Left panel: Band dispersion of the effective 3-orbital Hamiltonian
(solid line) and total LMTO band structure (dashed line) of LiRh$_2$O$_4$.
Right panel: partial $a_{1g}$ and $e_g^\pi$ density of states for the model.
LDA [reproduced from \cite{Arita08}].
}}
\label{Fig:LDA}
\end{figure}

Starting from the experimental crystal structure Fig.\ \ref{Fig:Crystal},
the first LDA+DMFT step is the calculation of the LDA bandstructure.
Our results, using linearized muffin tin orbnitals (LMTOs) \cite{LMTO}, are
shown in Fig. \ref{Fig:LDA} (left panel; dashed line).
We further simplified the
LDA bandstructure by a Wannier projection\cite{Vanderbilt} of the LMTO wave funtions
onto the subspace of Bloch waves, which were in turn Fourier transformed 
to Wannier funtions, see \cite{Projection} for details.
Here,  we even model the
LDA bandstructure by a two-band model (solid line).

\begin{figure}[tb]
\begin{center}
\includegraphics[width=10.5cm]{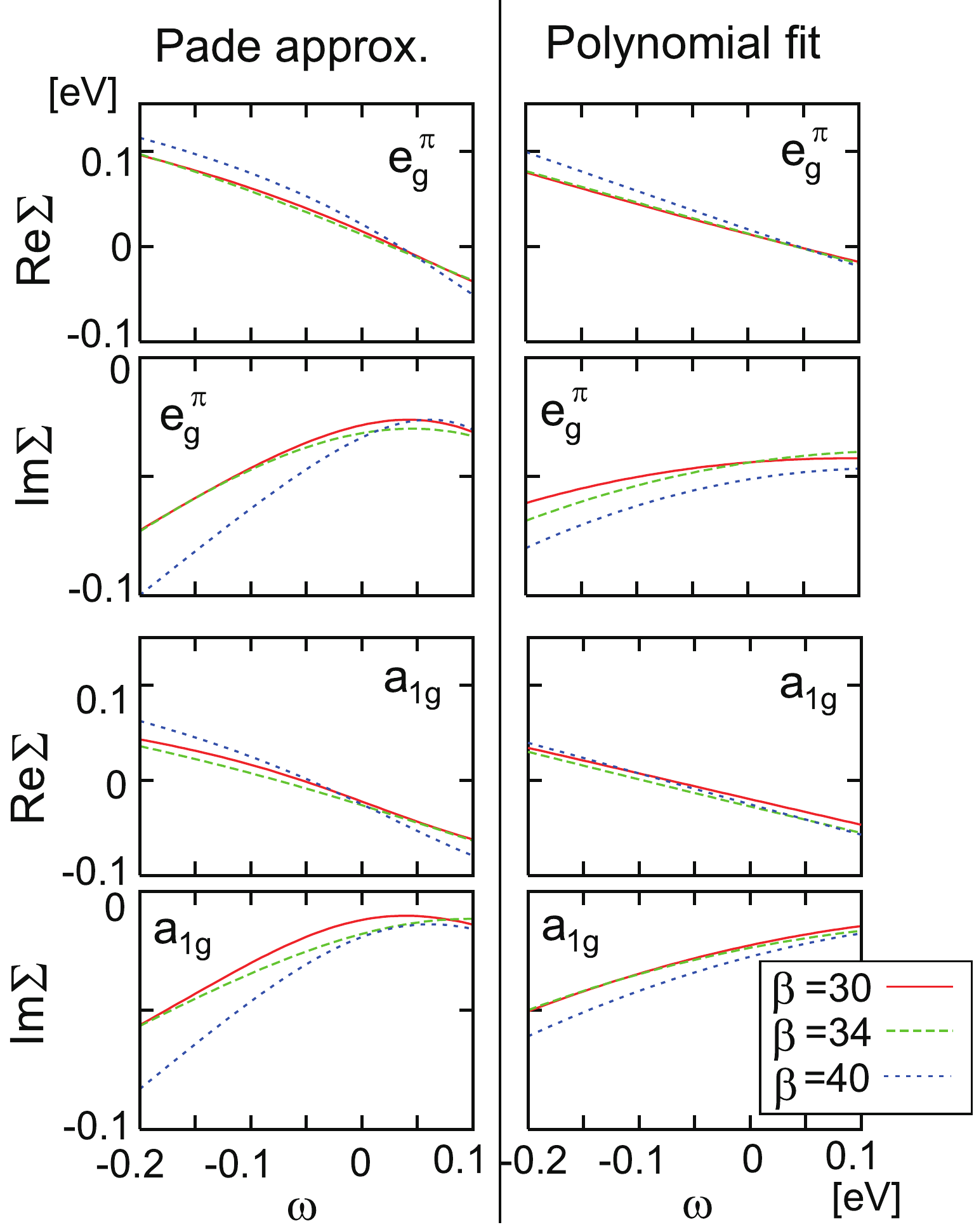}
\end{center}
\caption{DMFT(QMC)
self energy calculated by the Pad\'e approximation (left)
and a polynomial fit (right) [reproduced from \cite{Arita08}].
}
\label{Fig:Sigma}
\end{figure}

The next step is a self-consistent DMFT calculation. To this end, quantum Monte-Carlo simulations were used as an impurity solver \cite{Hirsch}.
The Coulomb interaction parameters were estimated as 
$(U, U',J)=(3.1, 1.7, 0.7)$eV   from \cite{Pchelkina} and temperatures
$\beta=1/k_BT=$30, 34, 40 eV$^{-1}$ were considered. 
From the imaginary QMC self energy we obtained the
 self energy on the real axis (Fig.\ \ref{Fig:Sigma})
 through a Pad\'e and polynomial (Taylor) fit. As one can see from
a comparison of the two fits there is some uncertainty, but absolute
differences are small, i.e., of $O(0.01)$ eV. Nonetheless, 
we proceeded with both self energies to have an estimate of the error.

From the self energy, Fig.\ \ref{Fig:Sigma}, we can estimate the 
quasiparticle weight $Z=(1-\partial {\rm Re}\Sigma/\partial \omega)^{-1}$ and the
effective mass enhancement $m^*/m=1/Z$. This effective mass enhancement is actually not very strong, i.e.,  $\approx 40\%$ for the $e_g^\pi$ band and
 $\approx 30\%$ for the $a_{1g}$ band. This indicates that electronic correlations are only intermediately strong for this compound, even though it is a transition metal oxide. The reason for this is the mixed-valent nature of
LiRh$_2$O$_4$ which puts the orbital occupation far away from a (more
strongly correlated) integer filling. A second noteworthy aspect,
we can extract from the self energy is the strong frequency dependence and
asymmetry of the imaginary part of the self energy. This poses the question
whether a constant-${\rm Im} \Sigma$, i.e., a constant relaxation time $\tau$ approach as in the much less involved Boltzmann approach, works.

From the (two) self energy of Fig.\ \ref{Fig:Sigma}, we calculated
the Seebeck coefficient using the formulas of Sec.\ \ref{sec:Post}.
As one can see there are some differences in the Seebeck coefficient for
the Pad\'e ($\times$ symbol) and polynomial fit ($*$ symbol), giving us an estimate of the accuracy of 
our calculation. Both are in 
good agreement with the experimental values \cite{Okamoto}.

\begin{figure}[t]
\begin{center}
\includegraphics[width=7.8cm]{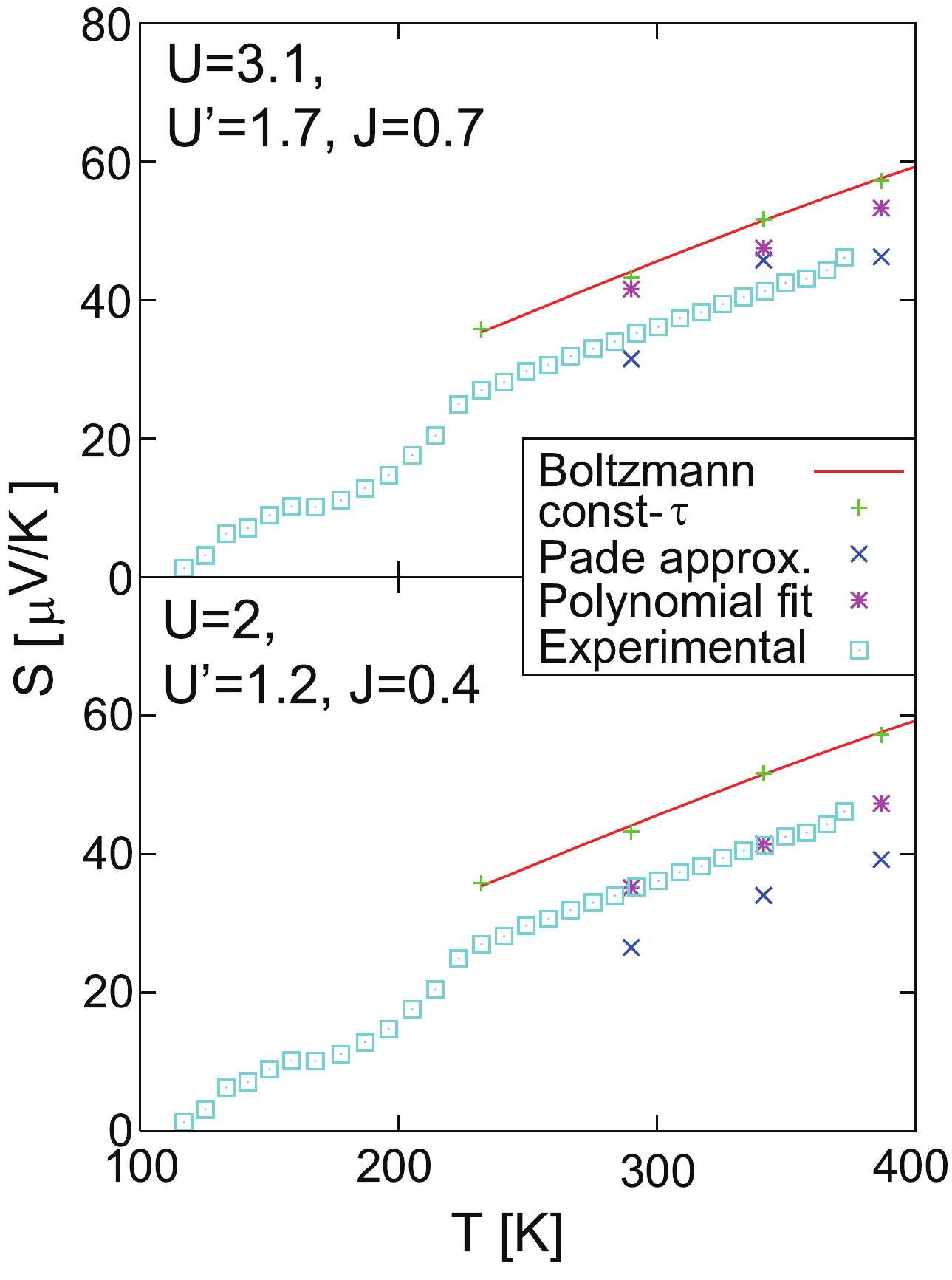}
\end{center}
\caption{
Thermopower calculated by 
the Boltzmann equation approach and the constant-$\tau$ method 
as well as by LDA+DMFT, using both
the Pad\'e approximation and a polynomial fit for the self energy [reproduced from \cite{Arita08}].
}
\label{Fig:S}
\end{figure}

Besides the LDA+DMFT study, we also performed  calculations (i) putting
a constant-$\tau$ self energy into the equations of Sec.\ \ref{sec:Post} ($+$ symbol) 
 and (ii) using directly the  Boltzmann equation (solid line).
As one can see in Fig.\ \ref{Fig:Sigma} both  agree, as one can expect
from theoretical considerations; but it is a good test in an actual 
implementation as two completely different programs based on
 different equations were employed. The Boltzmann equation yields a slightly too large Seebeck coefficient $S$, albeit it still agrees surprisingly well
with experiment. The reason for this is that  LiRh$_2$O$_4$ is
not strongly correlated. Besides, the two $e_g^\pi$ and $a_{1g}$ bands 
are not strongly shifted with respect to each other by electronic correlations
and have a not too different self energy. Hence, we are not too far from 
a situation
were the self energy is orbital independent. In this case, the DMFT spectral function is just a  more narrow (quasiparticle renormalized) version of the LDA DOS with the same height at the Fermi level. Because of this, relatively weak electronic correlations do not
strongly affect the Seebeck coefficient. We hence attribute the differences 
between Boltzmann approach and LDA+DMFT to the non-constant and strongly asymmetric
${\rm Im}\Sigma$.
This means that, in contrast to the constant-$\tau$ approximation,  
the actual life time of quasi-holes
is longer than that for quasi-particles.
 Let us emphasize that electronic correlations
play a much more prominent role in other transition metal oxides, so 
that for these the Boltzmann approach will fail.

\begin{figure}[tb]
\begin{center}
\includegraphics[width=8.5cm]{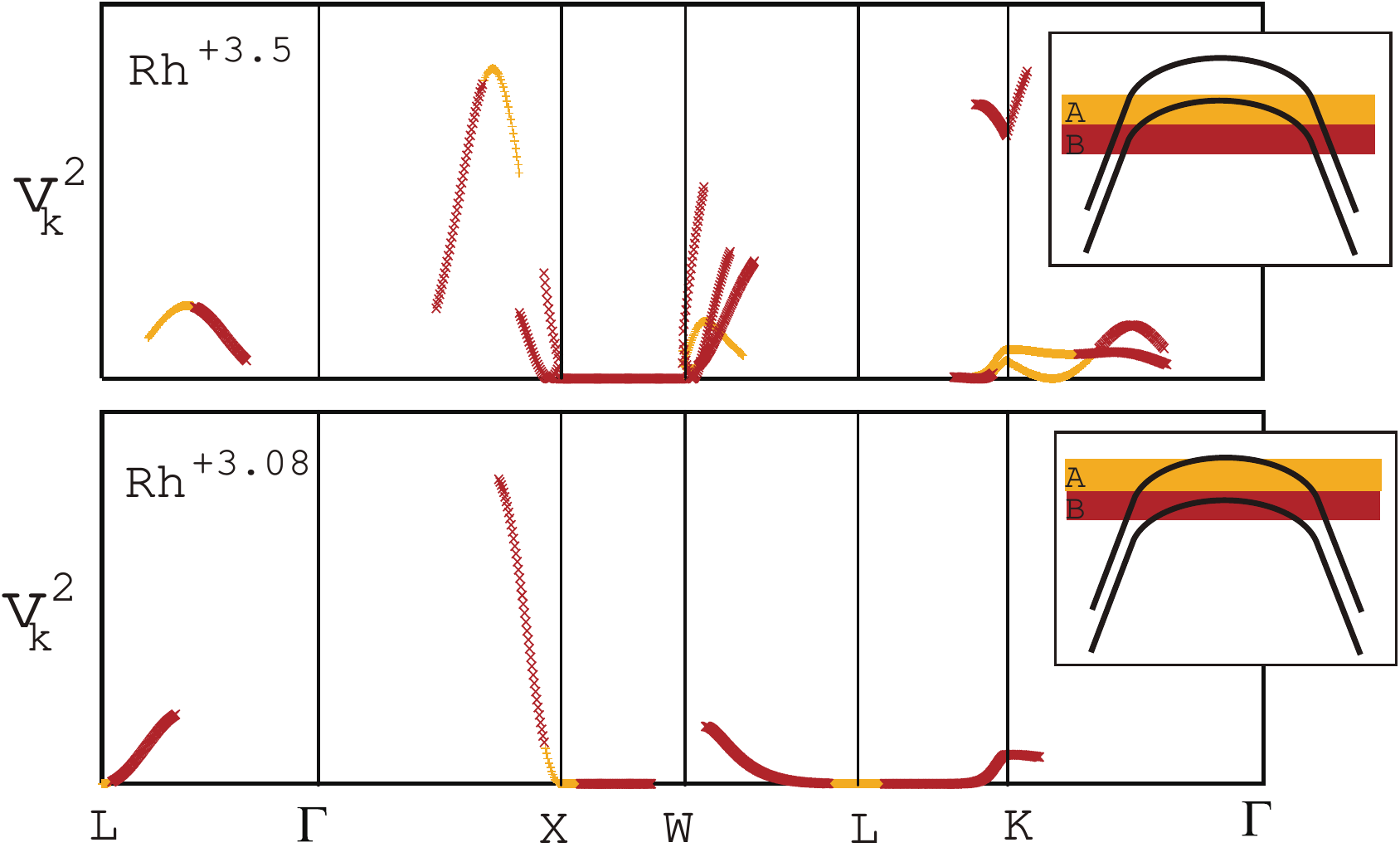}
\end{center}
\caption{
Group velocity squared ($v_k^2$) along different 
directions of the first Brillouin zone for Rh$^{+3.5}$ (LiRh$_2$O$_4$; upper panel) and Rh$^{+3.08}$ (electron-doped LiRh$_2$O$_4$; lower panel). 
$k$ point above the Fermi energy $E_F$
are shown in yellow, those below $E_F$ in red (reproduced from \cite{Arita08}).
}
\label{Fig:vel2all}
\end{figure}

Being confident, that the Boltzmann approach roughly describes the Seebeck coefficient of LiRh$_2$O$_4$, we analyze Eqs.\ (\ref{Eq:BA0}).
To this end, we plot in Fig.\ \ref{Fig:vel2all} the group
velocity along the indicated paths thought he Brillouin zone,
within the energy window of $|\varepsilon-E_F|<3k_BT$ at $T\simeq 300$K.
Fig.\ \ref{Fig:vel2all} (upper panel) shows that $v_B^2$ is  considerably larger than $v_B^2$ in large parts of the Brillouin zone, particularly
around the K and W point.
The reason for this difference is a particular shape of the 
bandstructure very similar to the ideas proposed in \cite{Kuroki}
for Na$_x$CoO$_2$. This pudding-mold type of shape is sketched
in the inset of  Fig.\ \ref{Fig:vel2all}, for the full bandstructure see
 Fig.\ \ref{Fig:LDA}. In contrast to  the one band situation in
 Na$_x$CoO$_2$ \cite{Kuroki}, we have however a double
pudding mold.
For the lower band, the pudding-mold shape leads to a very flat bandstructure
above the Fermi level, flatter than a simple maximum because
of additional turning points and  minima.
Consequently the group velocity above the Fermi level is very small and the Seebeck coefficient largely positive.

Having understood the bandstructure origin of the large thermopower in
 LiRh$_2$O$_4$, we are now in a position to identify routes to even further increase the
thermopower. As the sketch in  Fig.\ \ref{Fig:vel2all}
suggests the upper pudding mold band does not strongly contribute 
to the Seebeck coefficient or thermopower. Here, the situation is more like in a standard metal with
group velocities being large above {\em and} below the
Fermi level (this is the region   between $\Gamma$ and X and  between $\Gamma$ and L  point in the main panel). Therefore, positive and negative contributions
to the Seebeck coefficient roughly cancel for this upper pudding mold  band.

We can improve the situation however by electron doping which shifts the Fermi level to higher energy. Then the situation becomes very much the same as for the lower  pudding-mold band before and, at the same time, the lower  pudding-mold band is still contributing with the same sign because there are states below the Fermi level but no states above. As one can see in the lower main panel of  Fig.\ \ref{Fig:vel2all},  doping by 0.42 electrons, i.e., for a valence Rh$^{+3.08}$, indeed leads to a situation where only the squared group velocity below
the Fermi energy is large.

We further studied this idea by calculating the thermopower and
the power factor for various
Rh valences, using the Boltzmann equation approach. Note that we assumed
the electron doping not to affect the LDA bandstructure expect for a shift
of the Fermi level. We also neglected the energy  and filling dependence  
of $\tau$, which should be present and affect $\rho$
and, hence, the power factor (albeit not $S$).
Depending on how the electron-doping is realized $\tau$ might
change because of disorder effects.
Nonetheless,
we expect the tendencies to hold also for the power factor
in experiments electron-doping LiRh$_2$O$_4$.

\begin{figure}[tb]
\begin{center}
\vspace{.2cm}
\includegraphics[width=8.5cm]{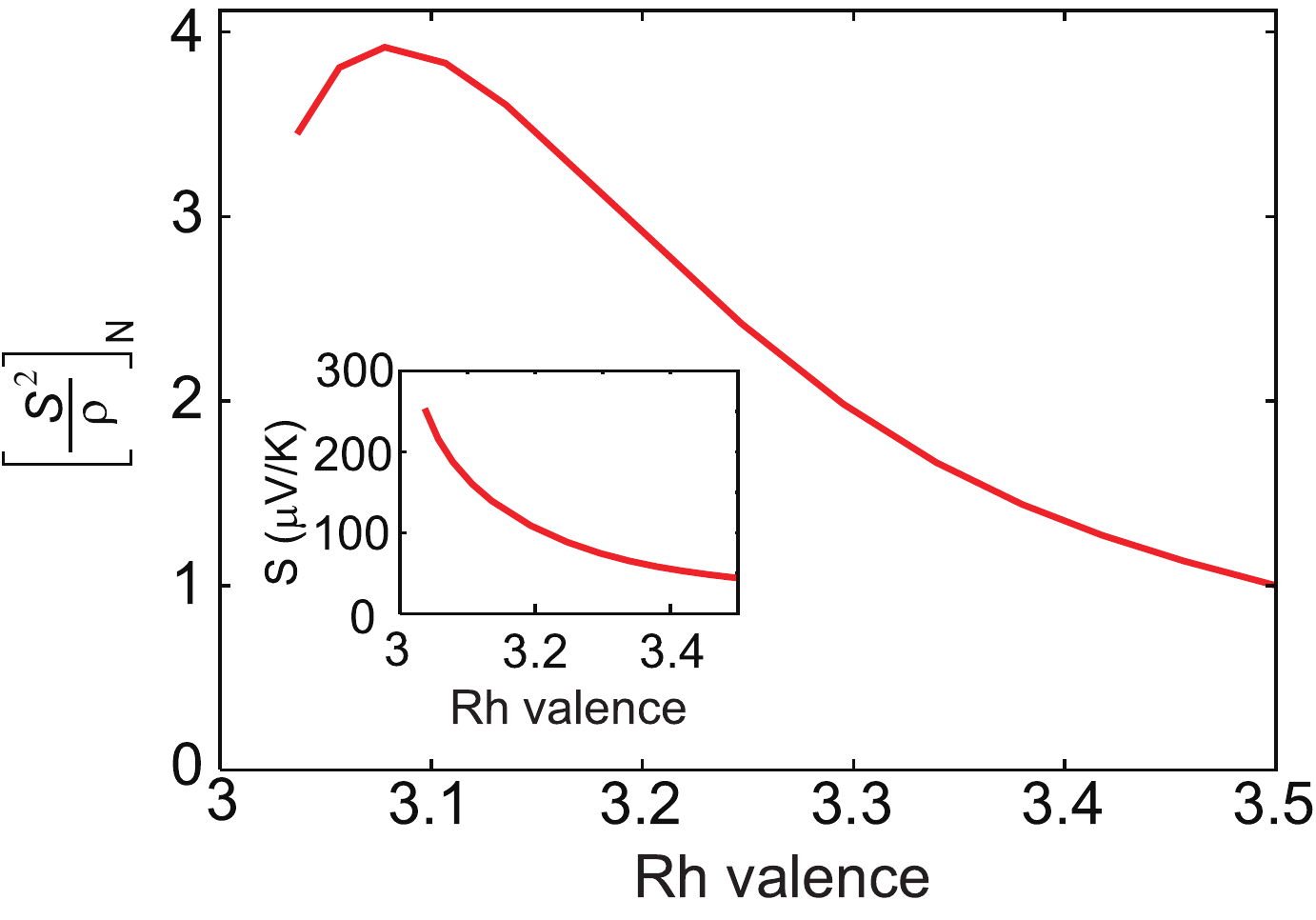}
\end{center}
\caption{(Color online)
Power factor (normalized by its value at Rh valence=+3.5 )
and thermopower (inset) as a function
of the valence of Rh, calculated by the Boltzmann equation. 
}
\label{Fig:val-seeb-pwf0026}
\end{figure}

As one can see in  Fig.\ \ref{Fig:val-seeb-pwf0026}
(inset) the Seebeck coefficient strongly increases with electron doping,
i.e., with reducing the Rh valence towards $3$.
However if the Rh valence is 3, the band is completely occupied so that 
while the Seebeck coefficient is large, the conductivity $\sigma=1/\rho$ becomes
small. Hence for the power factor
$S^2\sigma$  (Fig.\ \ref{Fig:val-seeb-pwf0026} main panel),  there is a trade-off
between a larger Seebeck factor and a large resistivity if the valence goes towards 3. This trade-off leads to a maximum for the
valence Rh$^{+3.08}$, afore shown in Fig.\ \ref{Fig:vel2all}.

\subsection{Summary and outlook}
\label{sec:Outlook}

We presented a brief introduction to the LDA+DMFT
approach for the realistic calculation of thermoelectric properties,
including bandstructure {\em and} electronic correlation effects.
We have shown that the LDA+DMFT 
results for LiRh$_2$O$_4$ well agree with experiment. Furthermore,
we identified the origin of the large thermopower in this material
to be a particular shape of the bandstructure of the form of
a (double) pudding mold. Even larger thermopowers can be obtained
if the material is electron-doped,
according to our prediction.
For the particular material  LiRh$_2$O$_4$
the microscopic mechanism for the large 
thermopower is foremost the bandstructure since electronic correlations
are not very strong (the effective mass enhancement is only 40\% and even
less for the $a_{1g}$ band). This shows the strength  of LDA+DMFT to unbiasedly identify 
bandstructure effects as the origin of large thermopowers where this
is appropriate and electronic correlations where these prevail.

For getting the optimal thermoelectric 
material, hetero- or nanostructure we likely need both ingredients.
First, a good bandstructure such as the pudding-mold form discussed in the
present paper which due to dramatically different group
velocities above and below the Fermi energy yields an extraordinarily large
Seebeck coefficient. And second, correlation effects which result in
 asymmetrical, sharply peaked renormalized spectra in the vicinity 
of the Fermi level which enhance the Seebeck coefficient as well.
With LDA+DMFT, we have an  ideal tool to scan and design
a wide range of potential SCES materials on a computer, providing experimental
physicists and chemists with valuable hints on how to improve the 
thermoelectric figure of merit. In the exemplary case of  LiRh$_2$O$_4$ this
would be through electron-doping the material.

\section*{Acknowledgment}
We would like to thank H. Takagi and Y. Okamoto 
for fruitful discussions;
numerical calculations were performed at
the facilities of the Supercomputer center,
ISSP, University of Tokyo.
This work was supported by Grants-in-Aid
for Scientific Research (MEXT Japan) grant
19019012,19014022, 19051016
and Russian Foundation for Basic Research (RFBR) grant 07-02-00041.

%
%

\end{document}